\documentclass[jcph]{apjrnl}
\usepackage{amsmath, bbm}
\usepackage[final]{graphicx}

\newcommand{\dnachd}[2]{\frac{\partial #1}{\partial #2}}

\newcommand{\tildeLapLong}{\tilde\Delta_L}

\begin{document}
\title{A multidomain spectral method for solving elliptic equations}

\author{Harald P. Pfeiffer\ref{add:a}, 
  Lawrence E. Kidder\ref{add:b},
  Mark A. Scheel\ref{add:d}, and 
  Saul A. Teukolsky\ref{add:c}
  \\
  \additem[add:a]{Department of Physics, Cornell University, Ithaca,
    New York 14853},
  \additem[add:b]{Center for Radiophysics and Space
    Research, Cornell University, Ithaca, New York 14853,}
  \additem[add:d]{California Institute of Technology, Pasadena,
    California 91125,} and
  \additem[add:c]{Department of Astrophysics,
    American Museum of Natural History, CPW \& 79th Street, New York,
    NY 10024. Permanent address: Department of Physics, Cornell
    University}
  \\
  E-mail: harald@astro.cornell.edu, kidder@astro.cornell.edu,
  scheel@tapir.caltech.edu, saul@astro.cornell.edu}

\date{February 27, 2002}
\maketitle

\begin{abstract}
  We present a new solver for coupled nonlinear elliptic partial
  differential equations (PDEs).  The solver is based on
  pseudo-spectral collocation with domain decomposition and can handle
  one- to three-dimensional problems.  It has three distinct features.
  First, the combined problem of solving the PDE, satisfying the
  boundary conditions, and matching between different subdomains is
  cast into one set of equations readily accessible to standard linear
  and nonlinear solvers.  Second, touching as well as overlapping
  subdomains are supported; both rectangular blocks with Chebyshev
  basis functions as well as spherical shells with an expansion in
  spherical harmonics are implemented. Third, the code is very
  flexible: The domain decomposition as well as the distribution of
  collocation points in each domain can be chosen at run time, and the
  solver is easily adaptable to new PDEs.  The code has been used to
  solve the equations of the initial value problem of general
  relativity and should be useful in many other problems.  We compare
  the new method to finite difference codes and find it superior in
  both runtime and accuracy, at least for the smooth problems
  considered here.
\end{abstract}


\section{Introduction}

Elliptic partial differential equations (PDE) are a basic and
important aspect of almost all areas of natural science.  Numerical
solutions of PDEs in three or more dimensions pose a formidable
problem, requiring a significant amount of memory and CPU time.
Off-the-shelf solvers are available; however, it can be difficult to
adapt such a solver to a particular problem at hand, especially when
the computational domain of the PDE is nontrivial, or when one deals
with a set of coupled PDEs.

There are three basic approaches to solving PDEs: Finite differences,
finite elements and spectral methods. Finite differences are easiest
to code. However, they converge only algebraically and therefore need
a large number of grid points and have correspondingly large memory
requirements.  Finite elements and spectral methods both expand the
solution in basis functions. Finite elements use many subdomains and
expand to low order in each subdomain, whereas spectral methods use
comparatively few subdomains with high expansion orders.  Finite
elements are particularly well suited to irregular geometries
appearing in many engineering applications.  For sufficiently regular
domains, however, spectral methods are generally faster and/or more
accurate.

Multidomain spectral methods date back at least to the work of
Orszag\cite{Orszag:1980}.  In a multidomain spectral method, one has
to match the solution across different subdomains.  Often this is
accomplished by a combination of solves on individual subdomains
together with a global scheme to find the function values at the
internal subdomain boundaries.  Examples of such global schemes are
relaxational iteration\cite{Funaro-Quarteroni-Zanolli:1988}, an
influence matrix\cite{Macaraeg-Streett:1986,Boyd:2001}, or the
spectral projection decomposition
method\cite{Gervasio-Ovtchinnikov-Quarteroni:1997}.  For simple PDEs
like the Helmholtz equation, fast solvers for the subdomain solves are
available. For more complicated PDEs, or for coupled PDEs, the
subdomain solves will typically use an iterative solver.  One drawback
of these schemes is that information from the iterative subdomain
solves is not used in the global matching procedure until the
subdomain solves have completely converged.
The question arises whether efficiency can be improved by avoiding
this delay in communication with the matching procedure.
\\

In this paper we present a new spectral method for coupled nonlinear
PDEs based on pseudospectral collocation with domain decomposition.
This method does not split subdomain solves and matching into two
distinct elements.  Instead it combines satisfying the PDE on each
subdomain, matching between subdomains, and satisfying the boundary
conditions into one set of equations.  This system of equations is
then solved with an iterative solver, typically GMRES\cite{Templates}.
At each iteration, this solver thus has up-to-date information about
residuals on the individual subdomains and about matching and thus can
make optimal use of all information.

The individual subdomains implemented are rectangular blocks {\em and}
spherical shells.  Whereas either rectangular blocks (see e.g.
\cite{Demaret-Deville:1991, Ku:1995, Pinelli-Vacca-Quarteroni:1997})
or spherical shells\cite{Grandclement-Bonazzola-et-al:2001} have been
employed before, we are not aware of work using both.  The code
supports an arbitrary number of blocks and shells that can touch each
other and/or overlap.

Moreover, the operator $\cal S$ at the core of the method (see section
\ref{sec:OperatorS}) turns out to be modular, i.e. the code fragments
used to evaluate the PDE, the boundary conditions, and the matching
conditions are independent of each other. Thus the structure of the
resulting code allows for great flexibility, which is further enhanced
by a novel point of view of the mappings that are used to map
collocation coordinates to the physical coordinates. This flexibility
is highlighted in the following aspects:
\begin{itemize}
\item The user code for the particular PDE at hand is completely
  independent from the code dealing with the spectral method and
  domain decomposition. For a new PDE, the user has to supply only the
  code that computes the residual and its linearization.
\item {\em Mappings} are employed to control how collocation points
  and thus resolution are distributed within each subdomain.  New
  mappings can be easily added which are then available for {\em all}
  PDEs that have been coded.
\item The solver uses standard software packages for the
  Newton-Raphson step, the iterative linear solvers, and the
  preconditioning.  Thus one can experiment with many different linear
  solvers and different preconditioners to find an efficient
  combination.  The code will also automatically benefit from
  improvements to these software packages.
\item The code is dimension independent (up to three dimensions).
\item Many properties of a particular solve can be chosen at runtime,
  for example the domain decomposition, the mappings used in each
  subdomain, as well as the choice of the iterative solver.  The user
  can also choose among differential operators and boundary conditions
  previously coded at runtime.
\end{itemize}

In the next section we recall some basics of the pseudo-spectral
collocation method. In section \ref{sec:Implementation} we describe
our novel approach of combining matching with solving of the PDE. For
ease of exposition, we interweave the new method with more practical
issues like mappings and code modularity. The central piece of our
method, the operator $\cal S$, is introduced in section
\ref{sec:OperatorS} for a one-dimensional problem and then extended to
higher dimensions and spherical shells.  In section \ref{sec:Examples}
we solve three example problems. Many practical issues like
preconditioning and parallelization are discussed in this section, and
we also include a detailed comparison to a finite difference code.

\section{Spectral Methods}
\label{sec:SpectralMethods}

We deal with a second order nonlinear elliptic partial differential
equation or system of equations,
\begin{equation}\label{eq:PDE}
  ({\cal N}u)(x)=0,\qquad x\in {\cal D},
\end{equation}
in some domain  ${\cal D}\subset\mathbbmss{R}^d$
with boundary conditions
\begin{equation}\label{eq:bc}
  g(u)(x)=0,\qquad x\in \partial{\cal D}.
\end{equation}
The function $u$ can be a single variable giving rise to a single PDE,
or it can be vector-valued giving rise to a coupled set of PDEs.
Throughout this paper we assume that the problem has a unique
solution.  We also absorb a possible right-hand side into the elliptic
operator $\cal N$.

The fundamental idea of spectral methods is to approximate the
solution to the PDE \eqnref{eq:PDE} as a series in some basis
functions $\Phi_k(x)$:
\begin{equation}
  \label{eq:Expansion}
  u(x) \approx u^{(N)}(x)\equiv \sum_{k=0}^{N}\tilde u_{k}\Phi_k(x).
\end{equation}
The coefficients $\tilde u_k$ are called the {\em spectral
  coefficients}.  The power of spectral methods stems from two simple
facts:
\begin{enumerate}
\item When approximating a smooth function with a series
  \eqnref{eq:Expansion}, the error of the approximation decreases {\em
    exponentially} with the number of basis functions $N$. Hence
  $u^{(N)}$ will converge toward the true solution $u$ of the PDE
  exponentially, provided $u$ is smooth and one can determine the
  spectral coefficients $\tilde u_k$ sufficiently well.
\item One can evaluate the action of $\cal N$ on the function
  $u^{(N)}$ {\em exactly} (i.e. up to numerical round-off).  This fact
  allows one to find the $\tilde u_k$ accurately.
\end{enumerate}

The second fact arises because the derivatives of the basis functions
are known analytically, and so by Eq.~\eqnref{eq:Expansion},
\begin{equation}
  \frac{du^{(N)}(x)}{dx}=\sum_{k=0}^N\tilde u_k\frac{d\Phi_k(x)}{dx},
\end{equation}
and similarly for higher derivatives.

In order to compute the spectral coefficients $\tilde u_k$ we use 
{\em pseudo-spectral collocation} where one requires
\begin{equation}\label{eq:PSC}
  ({\cal N}u^{(N)})(x_i)=0, \qquad i=0, \ldots, N.
\end{equation}
The points $x_i$ are called {\em collocation points}, and are chosen
as the abscissas of the Gaussian quadrature associated with the basis
set $\Phi_k$.  This choice of collocation points can be motivated by
considering $\int\left[ ({\cal N} u^{(N)})(x)\right]^2dx$.  Evaluating
this integral with Gaussian quadrature, we find by virtue of
Eq.~\eqnref{eq:PSC}
\begin{equation}
  \int_{\cal D}\left[\left({\cal N}u^{(N)}\right)(x)\right]^2\,dx
\approx\sum_{i=0}^Nw_i \left[\left({\cal N}u^{(N)}\right)(x_i)\right]^2=0,
\end{equation}
where $w_i$ are the weights of the quadrature.  We see that ${\cal
  N}u^{(N)}$ must be small throughout $\cal D$ and thus the function
$u^{(N)}$ satisfying Eqs.~\eqnref{eq:PSC} must be close to the true
solution. More rigorous treatments of the pseudospectral collocation
method can be found in the literature\cite{Gottlieb-Orszag:1977,
  Canuto-Hussaini, Boyd:2001}.

\subsection{Chebyshev polynomials}

Chebyshev polynomials are widely used as basis functions for spectral
methods. They satisfy the convenient analytical properties of
``classical'' orthogonal polynomials. Their defining differential
equation is a {\em singular} Sturm-Liouville problem, and so Chebyshev
expansions converge exponentially for smooth functions $u$ {\em
  independent} of the boundary conditions satisfied by $u$
\cite{Gottlieb-Orszag:1977, Boyd:2001}.

Chebyshev polynomials are defined by
\begin{equation}
  \label{eq:Chebyshev}
  T_k(X)=\cos(k \arccos X), \qquad X\in [-1,1].
\end{equation}
They are defined on the interval $X\in [-1,1]$ only; usually one needs
to map the {\em collocation coordinate} $X\in[-1,1]$ to the physical
coordinate of the problem, $x\in [a,b]$. We use the convention that
the variable $X$ varies over the interval $[-1,1]$, whereas $x$ is
defined over arbitrary intervals.  We will describe our approach to
mappings below in the implementation section.

For an expansion up to order $N$ (i.e. having a total of
$N+1$ basis functions) the associated collocation points are
\begin{equation}\label{eq:ChebyshevCollocationPoints}
  X_i=\cos\left(\frac{i\pi}{N}\right),\quad i=0, \ldots, N.
\end{equation}
Define the real space values 
\begin{equation}\label{eq:Chebyshev-Expansion}
  u_i\equiv u^{(N)}(X_i)=\sum_{k=0}^N\tilde u_k T_k(X_i).
\end{equation}
Using the discrete orthogonality relation
\begin{equation}
  \label{eq:Chebyshev-Orthogonality}
  \delta_{jk}=\frac{2}{N\bar c_k}\sum_{i=0}^N
  \frac{1}{\bar c_i}T_j(X_i)T_k(X_i)
\end{equation}
with 
\begin{equation}
  \bar c_i=\left\{\begin{array}{l}2\qquad k=0\mbox{ or }k=N\\
      1\qquad k=1,\ldots, N-1,\end{array}\right.
\end{equation}
we can invert \eqnref{eq:Chebyshev-Expansion} and find
\begin{equation}\label{eq:Chebyshev-inverseExpansion}
  \tilde u_j=\frac{2}{N\bar c_j}\sum_{i=0}^{N}\frac{u_i}{\bar c_i}T_j(X_i).
\end{equation}

Both matrix multiplications \eqnref{eq:Chebyshev-Expansion} and
\eqnref{eq:Chebyshev-inverseExpansion} can be performed with a fast
cosine transform in ${\cal O}(N\ln N)$ operations, another reason for
the popularity of Chebyshev basis functions. 

There are the same number of real space values $u_i$ and spectral
coefficients $\tilde u_k$, and there is a one-to-one mapping between
$\{u_i\}$ and $\{\tilde u_k\}$. Hence one can represent the function
$u^{(N)}$ by either $\{u_i\}$ or $\{\tilde u_k\}$.

The spectral coefficients of the derivative, 
\begin{equation}\label{eq:ExpansionDerivative}
  \frac{du^{(N)}}{dX}(X)=\sum_{k=0}^{N}\tilde u_k'T_k(X),
\end{equation}
are given by the recurrence relation
\begin{equation}\label{eq:DerivativeRecurrence}
\begin{aligned}
  \tilde u_i'&=\tilde u_{i+2}'+2(i+1)\tilde u_{i+1},\qquad i=1,\ldots,N-1,\\
  \tilde u_0'&=\frac{1}{2}\tilde u_2'+\tilde u_1,
\end{aligned}
\end{equation}
with $\tilde u_{N+1}=\tilde u_N=0$.  The coefficients of the second
derivative,
\begin{equation}
  \frac{d^2u^{(N)}}{dX^2}(X)=\sum_{k=0}^{N-1}\tilde u_k^{\prime\prime}T_k(X),
\end{equation}
are obtained by a similar recurrence relation, or by applying
\eqnref{eq:DerivativeRecurrence} twice.

\subsection{Basis functions in higher dimensions}
\label{sec:SpecMethods-Basisfunctions}

In higher dimensions one can choose tensor grids of lower dimensional
basis functions.  For example, a $d$-dimensional cube $[-1, 1]^d$ can
be described by Chebyshev polynomials along each coordinate axis.  For
a three-dimensional sphere or spherical shell, tensor products of
spherical harmonics for the angles and a Chebyshev series for the
radial coordinate\cite{Grandclement-Bonazzola-et-al:2001} are used.
It is also possible to expand the angular piece in a double
Fourier-series\cite{Orszag:1974}.

\subsection{Domain Decomposition}



If the computational domain $\cal D$ has a different topology than the
basis functions, then an expansion in the basis functions cannot cover
$\cal D$ completely. Moreover, the particular problem at hand might
require different resolution in different regions of the computational
domain which will render a single overall expansion inefficient.

One circumvents these problems with domain decomposition. The computational
domain $\cal D$ is covered by $N_{\cal D}$ subdomains 
\begin{equation}
  {\cal D}=\bigcup_{\mu=1}^{N_{\cal D}}{\cal D}_\mu,
\end{equation}
each having its own set of basis functions and expansion coefficients:
\begin{equation}
  u^{(\mu)}(x)=\sum_{k=0}^{N_\mu}\tilde u^{(\mu)}_k\Phi^{(\mu)}_k(x),
  \qquad x\in {\cal D}_\mu,\quad \mu=1, \ldots N_{\cal D}.
\end{equation}
Here $u^{(\mu)}$ denotes the approximation in the $\mu$-th domain, and
we have dropped the additional label $N$ denoting the expansion order
of $u^{(\mu)}$.  The individual subdomains ${\cal D}_\mu$ can touch
each other or overlap each other. To ensure that the functions
$u^{(\mu)}$ ---each defined on a single subdomain ${\cal D}_\mu$
only--- actually fit together and form a smooth solution of the PDE
\eqnref{eq:PDE} on the full domain ${\cal D}$, they have to satisfy
matching conditions.  In the limit of infinite resolution, we must
have that
\begin{itemize}
\item for touching subdomains ${\cal D}_\mu$ and ${\cal D}_\nu$ the
  function and its normal derivative must be smooth on the surface
  where the subdomains touch:
\begin{equation}
  \label{eq:Matching-touch}
\begin{aligned}
  u^{(\mu)}(x)&=u^{(\nu)}(x)\\
  \dnachd{u^{(\mu)}}{n}(x)&=-\dnachd{u^{(\nu)}}{n}(x)
  \end{aligned}
  \qquad
   x\in \partial{\cal D}_\mu\cap\partial{\cal D}_\nu
\end{equation}
(The minus sign in the second equation of \eqnref{eq:Matching-touch} 
occurs because we use the outward-pointing normal in each subdomain.)

\item for overlapping subdomains ${\cal D}_\mu$ and ${\cal D}_\nu$ the
  functions $u^{(\mu)}$ and $u^{(\nu)}$ must be identical in ${\cal
    D}_\mu\cap{\cal D}_\nu$. By uniqueness of the solution of the PDE,
  it suffices to require that the functions are identical on the
  boundary of the overlapping domain:
  \begin{equation}\label{eq:Matching-overlap}
  u^{(\mu)}(x)=u^{(\nu)}(x),
  \qquad x\in \partial\left({\cal D}_\mu\cap{\cal D}_\nu\right).
  \end{equation}

\end{itemize}

We will see in the next section how these conditions are actually
implemented in the code.

\section{Implementation} 
\label{sec:Implementation}

In this section we describe our specific approaches to several aspects
of multi-dimensional pseudo-spectral collocation with domain
decomposition.

\subsection{One-dimensional Mappings} 

Chebyshev polynomials are defined for $X\in [-1, 1]$.
Differential equations in general will be defined on a different
interval $x\in [a,b]$. In order to use Chebyshev polynomials, one
introduces a mapping
\begin{equation}
X: [a, b]\to [-1, 1],\quad x\to X=X(x)
\end{equation}
that maps the {\em physical coordinate} $x$ onto the {\em collocation
coordinate} $X$.

One could explicitly substitute this mapping into the PDE under
consideration. Derivatives would be multiplied by a Jacobian, and we would
obtain the PDE on the interval $[-1, 1]$.  For example, the
differential equation in the variable $x$
\begin{equation}\label{eq:Example}
  \dnachd{^2u}{x^2}+u=0,\qquad x\in [a,b],
\end{equation}
becomes the following differential equation in the variable $X$:
\begin{equation}\label{eq:ExampleMapped}
  {X'}^2\dnachd{^2u}{X^2}+X^{\prime\prime}\dnachd{u}{X}+u=0,\qquad X\in [-1,1],
\end{equation}
where $X'=\partial X/\partial x$ and $X^{\prime\prime}=\partial^2X/\partial x^2$.
Now one could expand $u(X)$ in Chebyshev polynomials, compute
derivatives $\partial/\partial X$ via the recurrence relation
\eqnref{eq:DerivativeRecurrence} and code
Eq.~\eqnref{eq:ExampleMapped} in terms of $\partial u/\partial X$.
This approach is common in the literature
\cite{Boyd:2001,Kidder-Finn:2000}. However, it has several
disadvantages: As one can already see from this simple example, the
equations become longer and one has to code and debug more terms.
Second, and more important, it is inflexible, since for each different
map one has to derive and code a mapped equation
\eqnref{eq:ExampleMapped}.  {\em A priori} one might not know the
appropriate map for a differential equation, and in order to try
several maps, one has to code the mapped equation several times.
Also, for domain decomposition, a different map is needed for each
subdomain.

We propose a different approach.  We still expand in terms of
Chebyshev polynomials on $X\in [-1, 1]$ and obtain the physical
solution via a mapping $X(x)$,
\begin{equation}
  u(x)=\sum_{k=0}^N\tilde u_kT_k(X(x)),
\end{equation}
and we still compute $\partial u(X)/\partial X$ and
$\partial^2u(X)/\partial X^2$ via the recurrence relation
\eqnref{eq:DerivativeRecurrence}. However, now we do {\em not}
substitute $\partial u(X)/\partial X$ and
$\partial^2u(X)/\partial X^2$ into the mapped differential
equation, Eq.~\eqnref{eq:ExampleMapped}.  Instead we compute first
numerically
\begin{align}\label{eq:FirstDeriv-Mapped}
  \dnachd{u(x)}{x}&=X'\dnachd{u(X)}{X}\\
\label{eq:SecondDeriv-Mapped}
  \dnachd{^2u(x)}{x^2}&
  ={X'}^2\dnachd{^2u(X)}{X^2}+X^{\prime\prime}\dnachd{u(X)}{X}
\end{align}
and substitute these values into the original physical differential
equation \eqnref{eq:Example}.  The collocation points are thus mapped
to the physical coordinates
\begin{equation}
  \label{eq:PhysicalGridPoints}
  x_i=X^{-1}(X_i).
\end{equation}

This approach separates the code into three distinct parts: 
\begin{enumerate}
\item Code dealing with the basis functions: transforms between
  collocation space $X$ and spectral space, evaluation of derivatives
  $\partial/\partial X$ via recurrence relations. This code depends
  only on the collocation coordinates $X\in [-1,1]$ (and on the
  angular coordinates $\theta, \phi$ for spherical shells).
\item Mappings that map between collocation coordinate $X$ and
  physical coordinates $x$.
\item The ``user code'' implementing the physical PDE [in our example
  Eq.~\eqnref{eq:Example}] that deals only with the physical coordinate
  $x$.
\end{enumerate}

These three elements are independent of each other:
\begin{itemize}
\item A user who wants to code another differential equation has
  only to write the code that evaluates the differential operator
  ${\cal N}$ in the physical space with physical derivatives. Then
  immediately all previously coded mappings are available for this
  new differential equation, as well as all basis functions.

\item In order to introduce a new mapping, one has to code only four
  functions, namely $X(x)$, its inverse $x(X)$, as well as the
  derivatives $X'(x)$ and $X^{\prime\prime}(x)$.  This new map can then be used 
  for any differential equation already coded or to be coded later.
  
\item In order to switch to a different set of basis functions, one
  has only to code the transforms and the recurrence relations for
  the derivatives. 
\end{itemize}

In practice we use three different mappings
\begin{equation}\label{eq:Mappings}
\begin{array}{ll}
  \mbox{linear:}& X(x) = Ax+B\\
  \mbox{log:} &   X(x) = A\log(Bx+C)\\
  \mbox{inverse:}& X(x)= \frac{A}{x}+B
\end{array}
\end{equation}
In each case the constants $A, B$ are chosen such that $[a, b]$ is
mapped to $[-1, 1]$.  The log mapping has one additional parameter
which is used to fine-tune the relative density of collocation points
at both ends of the interval $[a, b]$.  We show the effects of
different mappings in our first example in section \ref{sec:Example1}.

\subsection{Basis functions and Mappings in higher Dimensions} 

\subsubsection{Rectangular Blocks}

In order to expand in a $d$-dimensional rectangular block, 
\begin{equation}
{\cal D}=[a_1,
b_1]\times[a_2,b_2]\times\ldots\times[a_d, b_d],  
\end{equation}
we use a tensor product of Chebyshev polynomials with a 1-$d$ mapping
along each coordinate axis:
\begin{equation}\label{eq:Block-nD}
  u(x_1, \ldots, x_d)
  =\sum_{k_1=0}^{N_1}\sum_{k_2=0}^{N_2}\cdots\sum_{k_d=0}^{N_d}
    \tilde u_{k_1\cdots k_d}
T_{k_1}\!\!\left(X^{(1)}(x_1)\right)\cdots
T_{k_d}\!\left(X^{(d)}(x_d)\right).
\end{equation}
We use $d$ mappings
\begin{equation}
  X^{(l)}: [a_l,b_l]\to [-1, 1],\quad l=1, \ldots d,
\end{equation}
and the collocation points in physical space are the mapped
collocation points along each dimension,
\begin{equation}
  x_{i_1\cdots i_d}=\Big(x_{i_1}^{(1)}, \ldots, x_{i_d}^{(d)}\Big),
\end{equation}
where the coordinate along the $l$-th dimension $x^{(l)}_{i_l}$ is
given by Eq.~\eqnref{eq:PhysicalGridPoints} using $X^{(l)}$.

Note that such a $d$-dimensional rectangle has as many spectral
coefficients $\tilde u_{k_1\cdots k_d}$ as grid point values
$u_{i_1\ldots i_d}=u(x_{i_1}, \ldots, x_{i_d})$.  Therefore we can
equivalently solve for the spectral coefficients or the real space
values.  We will solve for the real space values $u_{i_1\ldots i_d}$.

\subsubsection{Spherical Shell}

In a spherical shell with inner and outer radii $0<R_1<R_2$ we use a
mapping for the radial coordinate.  A function $u(r,\theta,\phi)$ is
thus expanded as
\begin{equation}\label{eq:ExpansionSphere}
  u(r, \theta, \phi)
  =\sum_{k=0}^{N_r}\sum_{l=0}^L\sum_{m=-l}^l
    \tilde u_{klm}T_k(X(r))Y_{lm}(\theta,\phi),
\end{equation}
where real-valued spherical harmonics are used:
\begin{equation}\label{eq:Ylm}
  Y_{lm}(\theta, \phi)\equiv\left\{
\begin{array}{lll}
P_l^{m}(\cos\theta)\cos(m\phi), &&m\ge 0\\
P_l^{|m|}(\cos\theta)\sin(|m|\phi),&&m<0
\end{array}\right.
\end{equation}
$P_l^m(\cos\theta)$ are the associated Legendre polynomials.
Associating the $\sin$-terms with negative $m$ is not standard, but
eliminates the need to refer to two sets of spectral coefficients, one
each for the $\cos$-terms and the $\sin$-terms.  The radial mapping
$X:[R_1,R_2]\to[-1,1]$ can be any of the choices in
Eq.~\eqnref{eq:Mappings}.  The radial collocation points $r_i, i=0,
\dots, N_r$ are given by Eq.~\eqnref{eq:PhysicalGridPoints}.

For the angle $\phi$, Eq.~\eqnref{eq:Ylm} leads to a Fourier series
with equally spaced azimuthal collocation points
\begin{equation}
  \phi_i=\frac{2\pi i}{N_\phi},\qquad i=0, 1, \ldots, N_\phi-1.
\end{equation}
There is a total of $N_\theta=L+1$ angular collocation points
$\theta_i$, which are the abscissas of Gauss-Legendre integration.

We employ the software package Spherepack\cite{spherepack-home-page}
which provides routines to compute the collocation points, transforms
and angular derivatives.

In order to resolve the full Fourier series in $\phi$ up to $m=L$, one
needs $N_\phi\ge 2L+1$, since for $N_\phi=2L$, the term $\sin(L\phi)$
vanishes at all collocation points $\phi_i$.  We use $N_\phi=2(L+1)$
since FFTs are more efficient with an even number of points.

The expansion \eqnref{eq:ExpansionSphere} has a total of
$(N_r+1)(L+1)^2$ spectral coefficients but a total of $(N_r+1)
N_\theta N_\phi=2(N_r+1)(L+1)^2$ collocation points.  This means a
spherical shell has {\em more} collocation points than spectral
coefficients and the expansion~\eqnref{eq:ExpansionSphere}
approximates the grid point values in a least-square sense
only\cite{Swarztrauber:1979}.  Performing a spectral transform and its
inverse will thus project the grid point values into a subspace with
dimension equal to the number of spectral coefficients.
The implications of this fact for our code are discussed below in
section~\ref{sec:S-in-Spheres}.


\subsubsection{Representation of vectors and derivatives}

In both kinds of subdomains, rectangular blocks and spherical shells,
we expand the {\em Cartesian components} of vectors, and we compute
{\em Cartesian} derivatives, $\partial/\partial x, \partial/\partial
y, \partial/\partial z$.  These quantities are smooth everywhere, and
thus can be expanded in scalar spherical harmonics.  By contrast, the
spherical components of a vector field in a spherical shell are
discontinuous at the poles and cannnot be expanded in scalar spherical
harmonics. One would have to use e.g. vector spherical
harmonics\cite{Swarztrauber:1979, Swarztrauber:1981}.

For a spherical shell we provide an additional wrapper around the
basis functions and the radial mapping that transforms polar
derivatives $ \partial/\partial r, \partial/\partial\theta,
\partial/\partial\phi$ to Cartesian derivatives.  This involves
multiplications by sines and cosines of the angles $\theta,\phi$ which
can be performed grid point by grid point in real space.
Alternatively, it can be done spectrally by expressing, e.g.
$\sin\theta$ in spherical harmonics, and then using addition theorems
to reduce products of spherical harmonics to simple sums.  Carrying
out the transformation spectrally is slightly better in practice.

Representing vectors and derivatives in Cartesian coordinates in both
kinds of subdomains increases flexibility, too. We can use the {\em
  same} code to evaluate the residual in {\em both} kinds of
subdomains.

\subsection{The operator \protect\boldmath${\cal S}$}
\label{sec:OperatorS}

We now introduce the operator $\cal S$, the centerpiece of our method.
It combines the solution of the PDE, the boundary conditions and
matching between different subdomains.

\begin{figure}
\begin{centering}
  \includegraphics[scale=0.95]{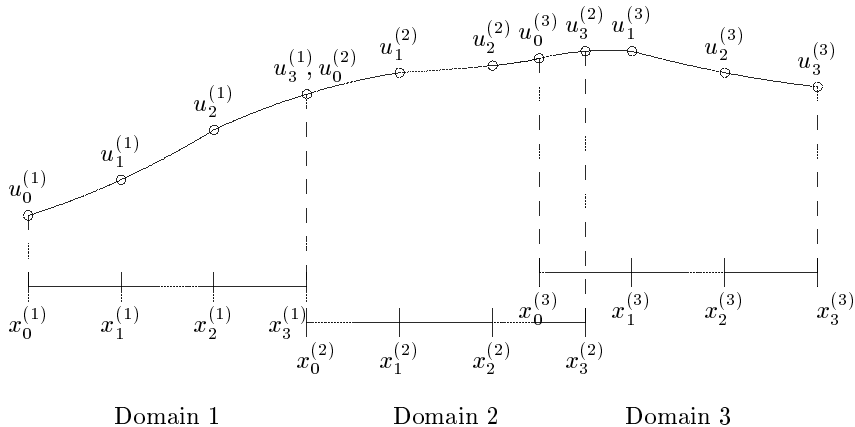}
\caption{\label{fig:Illustration}
  Illustration of matching with three subdomains in one dimension.
  Subdomains 1 and 2 touch each other, and subdomains 2 and 3 overlap.
  $x_i^{(\mu)}$ denotes the coordinate of the $i$-th collocation point
  of the $\mu$-th subdomain, and $u_i^{(\mu)}$ denotes the function
  values at the grid points.}
\end{centering}
\end{figure}
  
We introduce $\cal S$ first with a simple case, a one-dimensional
differential equation with a Dirichlet boundary condition at one end
and a von Neumann boundary condition at the other end:

\begin{align}
  ({\cal N}u)(x)&=0, \qquad a<x<b,\\
  u(a)&=A,\\
  \dnachd{u}{x}(b)&=B.
\end{align}
To explain our procedure for matching, we assume three domains as
depicted in Figure \ref{fig:Illustration}.  $N_\mu$ denotes the
highest retained expansion order in domain $\mu$; here
$N_\mu=3$ for all domains.  Domains 2 and 3 overlap.  Domains 1 and 2
touch so that the collocation points $x^{(1)}_3$ and $x^{(2)}_0$
represent the same physical point. The function value, however, is
represented twice, once assigned to domain 1, $u^{(1)}_3$, and once
belonging to domain 2: $u^{(2)}_0$. Using just the grid point values
within one subdomain, we can expand the function in that subdomain and
can evaluate derivatives. We can also interpolate the function to
arbitrary positions $x$. Thus, given values
$\{u^{(\mu)}_i, i=0, \ldots, N_\mu\}$, we can compute $u^{(\mu)}(x)$
for $x\in [x^{(\mu)}_0, x^{(\mu)}_{N_\mu}]$.

In order to determine the unknowns $u^{(\mu)}_i$, we need one equation
per unknown. We will first write down these equations and then explain
where they come from.
\begin{subequations}\label{eq:S3}
\begin{align}
  \label{eq:S3_A}
  \big({\cal N}u^{(\mu)}\big)\big(x^{(\mu)}_i\big)&=0,
  \quad  i=1,\ldots, N_\mu-1, \quad\mu=1,\ldots ,N_{\cal D}\\
  \label{eq:S3_B}
  u^{(1)}_0-A&=0\\
  \label{eq:S3_C}
  \dnachd{u^{(3)}}{x}\big(x^{(3)}_3\big)-B&=0\\
  \label{eq:S3_D}
  u^{(1)}_3-u^{(2)}_0&=0\\
  \label{eq:S3_E}
  \dnachd{u^{(1)}}{n}\big(x^{(1)}_3\big)
  +\dnachd{u^{(2)}}{n}\big(x^{(1)}_3\big)&=0\\
  \label{eq:S3_F}
  u^{(2)}_3 - u^{(3)}\big(x^{(2)}_3\big)&=0\\
  \label{eq:S3_G}
  u^{(3)}_0 - u^{(2)}\big(x^{(3)}_0\big)&=0
\end{align}
\end{subequations}
  
Eq.~\eqnref{eq:S3_A} represents the actual pseudo-spectral
equation~\eqnref{eq:PSC}.  It is enforced only for collocation points
that are {\em not} on the boundary of a subdomain.
Eqs.~\eqnref{eq:S3_B} and \eqnref{eq:S3_C} encode the boundary
conditions.  Eqs.~\eqnref{eq:S3_D} and \eqnref{eq:S3_E} describe the
value and derivative matching at touching subdomain boundaries. These
equations follow from Eq.~\eqnref{eq:Matching-touch}.
Eqs.~\eqnref{eq:S3_F} and \eqnref{eq:S3_G} perform matching between
overlapping subdomains as given by Eq.~\eqnref{eq:Matching-overlap}.

We will view the left-hand side of Eqs.~\eqnref{eq:S3} as a non-linear
operator $\cal S$. This operator acts on the set of grid point values
for {\em all} subdomains $\big\{u^{(\mu)}_i\big\}$ ($\mu=1,2,3,
i=0,\ldots, N_\mu$ in the example) and returns a residual that
incorporates the actual pseudo-spectral condition Eq.~\eqnref{eq:PSC},
the boundary conditions, and the matching conditions between different
subdomains.  If we denote the vector of {\em all} grid point values by
$\underline{\bf u}$, then the discretized version of the partial
differential equation becomes
\begin{equation}\label{eq:Su=0}
  {\cal S}\underline{\bf u}=0.
\end{equation}
The solution of Eq.~\eqnref{eq:Su=0} clearly is the solution of the
partial differential equation we want to obtain. By virtue of
Eq.~\eqnref{eq:Su=0} we thus have condensed the PDE, the boundary
conditions and matching into one set of nonlinear equations. 

We comment on some implementation issues:
\begin{itemize}
\item The action of the operator $\cal S$ can be computed very easily:
  Given grid point values $\underline{\bf u}$, every subdomain is
  transformed to spectral space and derivatives are computed.  Using
  the derivatives we can compute Eqs.~\eqnref{eq:S3_A},
  \eqnref{eq:S3_E} and any boundary conditions that involve
  derivatives like Eq.~\eqnref{eq:S3_C}. The interpolations necessary
  in Eqs.~\eqnref{eq:S3_F} and \eqnref{eq:S3_G} are done by summing
  up the spectral series.
  
\item $\cal S\underline{\bf u}$ can be computed in parallel:
  Everything except the matching conditions depends only on the set of
  grid point values within {\em one} subdomain. Therefore the natural
  parallelization is to distribute subdomains to different processors.
  
\item The code fragments implementing the nonlinear operator $\cal N$,
  the boundary conditions and the matching conditions are independent
  of each other. In order to change boundary conditions, one has
  only to modify the code implementing Eqs.~\eqnref{eq:S3_B} and
  \eqnref{eq:S3_C}.  In particular, the code for the matching-equations
  \eqnref{eq:S3_D}-\eqnref{eq:S3_G} can be used for {\em any}
  differential operator $\cal N$ and for {\em any} boundary condition.
\end{itemize}

We have now introduced the operator $\cal S$ in one dimension. Next we
address how to solve Eq.~\eqnref{eq:Su=0}, and then we generalize
$\cal S$ to higher dimensions. We present our method in this order
because the generalization to higher dimensions depends on some
details of the solution process.

\subsection{Solving \protect\boldmath${\cal S}u=0$}

In this section we describe how we solve the system of nonlinear
equations~\eqnref{eq:Su=0}.  Our procedure is completely standard and
requires three ingredients: A Newton-Raphson iteration to reduce the
nonlinear equations to a linear solve at each iteration, an iterative
linear solver, and the preconditioner for the linear solver.  For
these three steps we employ the software package
PETSc\cite{petsc-home-page}. We now comment on each of these three
stages.

\subsubsection{Newton-Raphson with line searches}

PETSc\cite{petsc-home-page} implements a Newton-Raphson method with
line searches, similar to the method described in
\cite{NumericalRecipes}.  Given a current guess $\underline{\bf
  u}_{\mbox{\footnotesize old}}$ of the solution, a Newton-Raphson
step proceeds as follows: Compute the residual
\begin{equation}
  \underline{\bf r}\equiv {\cal S}\underline{\bf u}_{\mbox{\footnotesize old}}
\end{equation}
and linearize $\cal S$ around the current guess $\underline{\bf
  u}_{\mbox{\footnotesize old}}$ of the solution:
\begin{equation}
  \label{eq:Jacobian}
  {\cal J}\equiv \frac{\partial\cal S}{\partial\underline{\bf u}}
(\underline{\bf u}_{\mbox{\footnotesize old}}).
\end{equation}
The Jacobian ${\cal J}$ is a $N_{DF}\times N_{DF}$-dimensional
matrix, $N_{DF}$ being the number of degrees of freedom. Next compute
a correction $\delta\underline{\bf u}$ by solving the linear system
\begin{equation}
\label{eq:Linearization}
 {\cal J}\delta\underline{\bf u}=-\underline{\bf r}. 
\end{equation}
Finally a line-search is performed in the direction of
$\delta\underline{\bf u}$. Parametrize the new solution by
\begin{equation}
  \underline{\bf u}_{\mbox{\footnotesize new}}
  =\underline{\bf u}_{\mbox{\footnotesize old}}
  +\lambda\;\,\delta\underline{\bf u}
\end{equation}
and determine the parameter $\lambda>0$ such that the residual of the
new solution,
\begin{equation}
  ||{\cal S}(\underline{\bf u}_{\mbox{\footnotesize new}})||,
\end{equation}
has sufficiently decreased. Of course, close enough to the true
solution, the full Newton-Raphson step $\lambda=1$ will lead to
quadratic convergence.  PETSc offers different algorithms to perform
this line-search.  The line search ensures that in each iteration the
residual does indeed decrease, which is not guaranteed in
Newton-Raphson without line searches.

\subsubsection{Linear Solve}

In each Newton-Raphson iteration one has to solve
Eq.~\eqnref{eq:Linearization}, a linear system of $N_{DF}$ equations.
For large systems of linear equations, iterative linear
solvers\cite{Templates} are most efficient. Such iterative solvers
require solely the ability to compute the matrix-vector product ${\cal
  J}\underline{\bf v}$ for a given vector $\underline{\bf v}$.  Since
spectral derivatives and spectral interpolation lead to {\em full}
(i.e. non-sparse) matrices it is impractical to set up the matrix
${\cal J}$ explicitly.  One can compute these matrix-vector products
instead with the linearized variant of the code that computes the
operator $\cal S$, i.e.  equations \eqnref{eq:S3_A}-\eqnref{eq:S3_G}
and their multidimensional generalizations.  Thus our method requires
the linearizations of the operator $\cal N$ [Eq.~\eqnref{eq:S3_A}] and
of the boundary conditions [Eqs.~\eqnref{eq:S3_B} and
\eqnref{eq:S3_C}].  The matching equations
\eqnref{eq:S3_D}-\eqnref{eq:S3_G} are linear anyway, so one can reuse
code from $\cal S$ for these equations.  The linearizations are merely
Frechet derivatives\cite{Boyd:2001} of the respective operators
evaluated at the collocation points, and therefore the Newton-Raphson
iteration applied to the discretized equations is equivalent to the
Newton-Kantorovitch iteration applied to the PDE.

PETSc includes several different linear iterative solvers (GMRES,
TFQR, ...)  that can be employed for the linear solve inside the
Newton-Raphson iteration.  The choice of linear solver and of options
for the linear solver and for the Newton-Raphson iteration are made at
runtime. This allows one to experiment with different linear solvers
and with a variety of options to find an efficient combination.  Note
that the matching conditions \eqnref{eq:Matching-touch} and
\eqnref{eq:Matching-overlap} lead to a nonsymmetric matrix $\cal J$.
Therefore only iterative methods that allow for nonsymmetric matrices
can be used.

\subsubsection{Preconditioning}\label{sec:Preconditioning}

In practice one will find that the Jacobian ${\cal J}$ is
ill-conditioned and thus the iterative method employed to solve
Eq.~\eqnref{eq:Linearization} will need an increasing number of
iterations as the number of collocation points is increased.  The {\em
  spectral condition number} $\kappa$ of a matrix is the ratio of
largest to smallest eigenvalue of this matrix,
\begin{equation}
  \kappa=\frac{\lambda_{max}}{\lambda_{min}}.
\end{equation}
For second order differential equations discretized with Chebyshev
polynomials, one finds $\kappa\propto N^4$, $N$ being the number of
grid points per dimension.  Solving a linear system to given accuracy
will require\cite{Axelsson:1994, Templates} ${\cal O}(\kappa)$ iterations of
the Richardson method, and ${\cal O}(\sqrt{\kappa})$ iterations of modern
iterative methods like conjugate gradients or GMRES.  Although modern
methods are better than Richardson iteration, it is still vital to
keep $\kappa$ close to $1$.

This is achieved with {\em preconditioning}. Instead of solving
Eq.~\eqnref{eq:Linearization} directly, one solves
\begin{equation}
 {\cal B}{\cal J}\;\delta\underline{\bf u}
=-{\cal B}\underline{\bf r},
\end{equation}
with the preconditioning matrix ${\cal B}$.  Now the iterative solver
deals with the matrix ${\cal B}\cal J$. If ${\cal B}$ is a good
approximation to ${\cal J}^{-1}$, then ${\cal B}{\cal J}$ will be
close to the identity matrix, the condition number will be close to
unity, and the linear solver will converge quickly. 

Hence the problem reduces to finding a matrix $\cal B$ that
approximates ${\cal J}^{-1}$ sufficiently well and that can be
computed efficiently.
There exist many different approaches, most notably finite difference
preconditioning\cite{Orszag:1980} and finite element
preconditioning\cite{Deville-Mund:1985}; we will follow a two-stage
process proposed by Orszag\cite{Orszag:1980}.  First, initialize a
matrix ${\cal A}_{FD}$ with a finite difference approximation of the
Jacobian $\cal J$.  Second, approximately invert ${\cal A}_{FD}$ to
construct ${\cal B}$,
\begin{equation}\label{eq:PC}
  {\cal B}\approx {\cal A}_{FD}^{-1}.
\end{equation}
In one spatial dimension ${\cal A}_{FD}$ is tridiagonal and direct
inversion ${\cal B}\equiv{\cal A}_{FD}^{-1}$ is feasible.  In two or
more dimensions, direct inversion of ${\cal A}_{FD}$ is too expensive;
for problems in one two-dimensional subdomain, hardcoded incomplete
LU-factorizations have been developed\cite{Canuto-Hussaini}.  In our
case we have to deal with the additional complexity that the Jacobian
and therefore ${\cal A}_{FD}$ contains matching conditions.  Since we
choose the domain decomposition at runtime, nothing is known about the
particular structure of the subdomains.

We proceed as follows: We initialize ${\cal A}_{FD}$ with the finite
difference approximation of ${\cal J}$.  It is sufficient to include
those terms of the Jacobian in ${\cal A}_{FD}$ that cause the
condition number to increase with the expansion order. These are the
second spatial derivatives and the first derivatives from matching
conditions and boundary conditions, Eqs.~\eqnref{eq:S3_E}
and~\eqnref{eq:S3_C}. Including the value matching conditions
\eqnref{eq:S3_D}, \eqnref{eq:S3_F}, \eqnref{eq:S3_G} in ${\cal
  A}_{FD}$ improves the ability of the preconditioner to represent
modes extending over several subdomains and thus decreases the number
of iterations, too.  In the first example in section
\ref{sec:Example1} we demonstrate that preconditioning is indeed
necessary, and that one should precondition not only the second order
derivatives, but also the matching conditions.  Some details about the
finite difference approximations are given in appendix
\ref{sec:FD-details}.

Having set up ${\cal A}_{FD}$ we then use the software package
PETSc\cite{petsc-home-page} for the approximate inversion of
Eq.~\eqnref{eq:PC}.  PETSc provides many general purpose
preconditioners that perform the step~\eqnref{eq:PC} either explicitly
or implicitly, most notably ILU and the overlapping Schwarz method.
With PETSc we can explore these to find the most efficient one.  We
will describe our particular choices for preconditioning below for
each example.

\subsection{\protect\boldmath${\cal S}$ in higher dimensions}

Generalizing $\cal S$ to multiple dimensions is conceptually
straightforward, since Eqs.~\eqnref{eq:S3} generalize nicely to higher
dimensions.  In order to simplify the matching between touching
subdomains, we require that on a surface shared by touching
subdomains, the collocation points are {\em identical}.  If, for
example, two three-dimensional rectangular blocks touch along the
$x$-axis, then both blocks must have identical lower and upper bounds
of the blocks along the $y$ and $z$ axis and both blocks must use the
same mappings and the same number of collocation points along the $y$-
and $z$-axis.  For concentric spherical shells, this restriction
implies that all concentric shells must have the same number of
collocation points in the angular directions.  With this restriction,
matching between touching subdomains remains a point-by-point
operation.

For overlapping domains, no restriction is needed. If a boundary point
of one subdomain happens to be within another subdomain, then an
equation analogous to \eqnref{eq:S3_F} is enforced using spectral
interpolation.

The actual implementation of the operator $\cal S$ involves
bookkeeping to keep track of which subdomains overlap or touch, or of
what equation to enforce at which grid point. When running in
parallel, matching conditions have to be communicated across
processors. We utilize the software package KeLP\cite{kelp-home-page},
which provides functionality to iterate over boundary points of a
specific subdomain.  It also provides routines to handle the
interprocessor communication needed for the matching conditions.

\subsection{Extension of \protect\boldmath$\cal S$ to Spherical Shells}
\label{sec:S-in-Spheres}

Spherical shells have the additional complexity of having more
collocation points than spectral coefficients, $N_{col}>N_{spec}$, at
least in our formulation.  Transforming to spectral space and back to
real space projects the real-space values into a
$N_{spec}$-dimensional subspace.  Since spectral transforms are used
for derivatives and interpolation, a sphere has effectively only
$N_{spec}$ degrees of freedom.  If we naively try to impose $N_{col}$
equations, one at each collocation point, and if we try to solve for
real space values at each collocation point, we find that the linear
solver does not converge.  This happens because more equations are
imposed than degrees of freedom are available. Thus we cannot solve
for the real space values $u_{ijk}$ in a spherical shell.

The next choice would be to solve for the spectral coefficients
$\tilde u_{klm}$ as defined in Eq.~\eqnref{eq:ExpansionSphere} This is
also problematic as it prohibits finite-difference preconditioning.
One finds guidance on how to proceed by considering the prototypical
elliptic operator, the Laplacian. Application of $\nabla^2$ to
an expansion in spherical harmonics yields
\begin{equation}
  \nabla^2\;\sum_{l,m}a_{lm}(r)Y_{lm}
=\sum_{l,m}\!\left[\!-\frac{l(l\!+\!1)a_{lm}(r)}{r^2}
  \!+\!\frac{1}{r^2}\dnachd{}{r}\!\left(\!r^2\dnachd{a_{lm}(r)}{r}\right)
\right]Y_{lm}.
\end{equation}
We see that the different $(lm)$-pairs are uncoupled. The angular
derivatives will therefore be diagonal in spectral space (with
diagonal elements $-l(l+1)/r^2$).  However, one has to precondition
the radial derivatives in order to keep the spectral conditioning
number low and must therefore keep real-space information about the
radial direction. We therefore solve for the coefficients $\hat
u_{ilm}$ of an expansion defined by
\begin{equation}\label{eq:ExpansionSphere2}
  u(r_i, \theta, \phi)
  =\sum_{l=0}^L\sum_{m=-l}^l
    \hat u_{ilm}Y_{lm}(\theta, \phi).
\end{equation}
This mixed real/spectral expansion has $N_{spec}$ coefficients $\hat
u_{ilm}$ and retains real space information about the radial
coordinate necessary for finite difference preconditioning.  In order
to precondition the flat space Laplacian in a spherical shell, ${\cal
  A}_{FD}$ is initialized with the diagonal matrix
$\mbox{diag}(-l(l+1)/r_i^2)$ for the angular piece of $\nabla^2$ and
with finite differences for the radial derivatives.  More general
differential operators are discussed in the last example, 
section~\ref{sec:Example3}, and in appendix
\ref{sec:Nonflat-Preconditioning}.

In order to evaluate ${\cal S}\underline{\bf u}$ for a spherical
shell, we proceed as follows.  $\underline{\bf u}$ contains the
coefficients $\hat u_{ilm}$.  Transform these coefficients to real
space values.  This involves only an angular transform.  Compute
boundary conditions, matching conditions, and the residual of the
nonlinear elliptic operator $\cal N$ at each collocation point as in
rectangular blocks.  At this stage we have $N_{col}$ collocation point
values, all of which should vanish for the desired solution. We
transform these values back into the coefficients of
Eq.~\eqnref{eq:ExpansionSphere2} and return these coefficients as the
residual of the operator $\cal S$.

\section{Examples} 
\label{sec:Examples}

\subsection{\protect\boldmath $\nabla^2u=0$ in 2-D}
\label{sec:Example1}
\begin{figure}
\begin{centering}
 \includegraphics[scale=0.5]{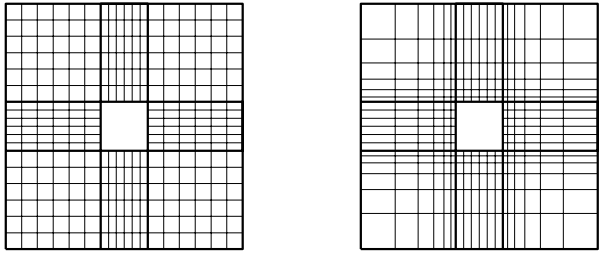}
\caption{\label{fig:SketchRectangles}Domain decomposition for Laplace 
  equation in a square. The left plot illustrates linear mappings in
  all subdomains and the right plot shows log-linear-log mappings along
  each axis.}
\end{centering}
\end{figure}

As a first test, we solve the Laplace equation in two dimensions with
Dirichlet boundary conditions:
\begin{align}\label{eq:Test1a}
  \nabla^2 u(x,y)&=0 &&(x,y)\in {\cal D}\\
  \label{eq:Test1b}
  u(x,y)&=\ln(x^2+y^2)&&(x,y)\in \partial{\cal D}
\end{align}
The computational domain ${\cal D}$ is a square with side $2L$
centered on the origin with a smaller square with side $2$ excised:
\begin{equation}
    {\cal D}=\{(x,y) | -L\le x,y\le L\} - \{(x,y)| -1<x,y<1\}
\end{equation}
This domain is decomposed into 8 touching rectangles as shown in
figure \ref{fig:SketchRectangles}. This figure also illustrates the
difference between linear mappings and logarithmic mappings. The right
plot of figure \ref{fig:SketchRectangles} shows that logarithmic
mappings move grid points closer to the excised rectangle.  For
clarity, both plots neglect the fact that the Chebyshev collocation
points given in Eq.~\eqnref{eq:ChebyshevCollocationPoints} are
clustered toward the boundaries.

\begin{figure}
\begin{centering}
\includegraphics[scale=0.4]{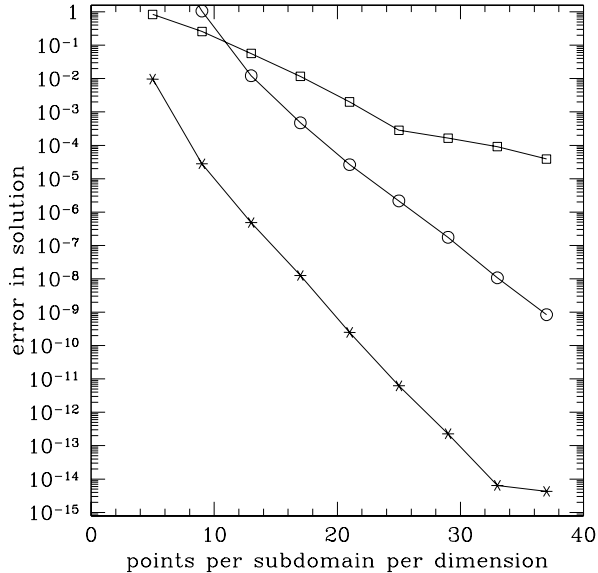}
\caption{\label{fig:Laplace2D}rms-errors of solution of Laplace 
  equation in a square. Stars denote $L=5$ with linear mappings,
  squares $L=100$ with linear mappings, and circles $L=100$ with
  log mappings.}
\end{centering}
\end{figure}

We solve Eqs.~\eqnref{eq:Test1a} and~\eqnref{eq:Test1b} for three
cases:
\begin{itemize}
\item $L=5$ with linear mappings
\item $L=100$ with linear mappings
\item $L=100$ with logarithmic mappings.
\end{itemize}
Equation~\eqnref{eq:Test1a} is linear, therefore only one
Newton-Raphson iteration with one linear solve is necessary.  The
numerical solution is compared to the analytic solution
$u(x,y)=\ln(x^2+y^2)$.  The errors are shown in figure
\ref{fig:Laplace2D}. In the small computational domain extending only
to $x,y=\pm 5$, the accuracy of the solution quickly reaches numerical
roundoff.  In the larger domain extending to $L=100$ the achievable
resolution with the same number of collocation points is of course
lower.  With linear mappings we achieve an absolute accuracy of
$10^{-4}$ with a total of $\sim 10000$ collocation points. This is
already better than finite difference codes. However this accuracy can
be increased with better adapted mappings. Since the solution
$\ln(x^2+y^2)$ changes much faster close to the origin than far away,
one expects better convergence if more collocation points are placed
close to the origin. This can be achieved with logarithmic mappings.
Figure \ref{fig:Laplace2D} shows that logarithmic mappings roughly
double the convergence rate. At the highest resolution the difference
is over four orders of magnitude.

\begin{table}
\begin{centering}
\caption{\label{tab:Laplace2D}
  Number of iterations $N_{its}$ in the linear solver for various 
  kinds of preconditioning as a function of the number of collocation 
  points $N$ per dimension per subdomain. {\em no PC} stands for no 
  preconditioning at all, 
  {\em PC $\nabla^2$} includes only second derivative terms in 
  ${\cal A}_{FD}$, whereas {\em full PC} includes also matching conditions.  
  ILU(2) decomposition, exact and approximate  inversion of 
  ${\cal A}_{FD}$ are explored.
}
 \begin{tabular}{|c|c|c|ccc|ccc|c|}\hline
   & &\multicolumn{4}{|c|}{${\cal B}=\mbox{ILU(2)}[{\cal A}_{FD}]$} & \multicolumn{3}{|c|}{ ${\cal B}={\cal A}_{FD}^{-1}$} & ${\cal B}\approx {\cal A}_{FD}^{-1}$ \\
    $N$ & no PC & PC $\nabla^2$ & \multicolumn{3}{|c|}{full PC} & \multicolumn{3}{|c|}{full PC} & full PC \\
        & $N_{its}$ & $N_{its}$ & $N_{its}$ & $\lambda_{max}$ & $\lambda_{min}$ & $N_{its}$ & $\lambda_{max}$ & $\lambda_{min}$ & $N_{its}$ \\\hline
    5 &    54 &    22 &  7 & 2.1 & 0.96 & 4 & 1.7 & 1.00 & 4 \\
    9 &   194 &    38 & 10 & 2.2 & 0.64 & 6 & 2.0 & 1.00 & 6 \\
   13 &   302 &    49 & 11 & 2.3 & 0.35 & 6 & 2.2 & 1.00 & 6 \\
   17 &   594 &    64 & 14 & 2.4 & 0.20 & 6 & 2.2 & 1.00 & 6 \\
   21 &   967 &   109 & 17 & 2.4 & 0.13 & 6 & 2.4 & 1.00 & 6 \\
   25 &  1244 &   140 & 20 & 2.5 & 0.09 & 6 & 2.4 & 1.00 & 7 \\
   29 &       &   206 & 25 & 2.6 & 0.05 & 6 & 2.5 & 1.00 & 7 \\ 
   33 &       &   255 & 27 & 2.6 & 0.04 & 7 & 2.4 & 1.00 & 7 \\\hline
  \end{tabular}
\end{centering}
\end{table}

Table \ref{tab:Laplace2D} compares the number of iterations $N_{its}$
in the linear solver for different choices of the finite difference
preconditioning matrix ${\cal A}_{FD}$ [section
\ref{sec:Preconditioning}]. Without any preconditioning, ${\cal
  A}_{FD}=\mathbf{1}$, $N_{its}$ increases very quickly with the
number of collocation points.  If only second derivative terms are
included in ${\cal A}_{FD}$ then $N_{its}$ grows more slowly. The
number of iterations becomes almost independent of resolution if both
second derivatives and the matching conditions \eqnref{eq:S3_D} and
\eqnref{eq:S3_E} are incorporated into ${\cal A}_{FD}$.  In this case
ILU(2) preconditioning completely controls the largest eigenvalue
$\lambda_{max}$, and it is then the smallest eigenvalue
$\lambda_{min}$ that causes the number of iterations to increase. It
is typical that ILU has difficulties controlling the long wavelength
modes, and the problem is aggravated because the subdomains are only
weakly coupled. Table \ref{tab:Laplace2D} also contains results for
exact inversion of ${\cal A}_{FD}$. Exact inversion controls
$\lambda_{min}$, too, and now the number of iterations is independent
of resolution. However, direct inversion is an expensive process and
only possible for small problems like this one. Finally, the table
also contains the iteration count for approximate inversion of ${\cal
  A}_{FD}$, which is our preferred method for more complex geometries
in 3 dimensions. It will be explained in detail in the next example.
The eigenvalues in table \ref{tab:Laplace2D} are estimates obtained by
PETSc during the linear solve.

\subsection{Quasilinear Laplace equation with two excised spheres}

This solver was developed primarily for elliptic problems in numerical
relativity. Accordingly we now solve an equation that has been of
considerable interest in that field over the last few years (see e.g.
\cite{Cook-Choptuik-etal:1993, Cook:2000} and references therein).
Readers not familiar with relativity can simply view this problem as
another test example for our new solver\footnote{The solution of this
  problem describes two black holes. The surfaces of the spheres
  $S_{1,2}$ correspond to the horizons of the black holes, the
  function $A^2$ encodes information about spins and velocities of the
  black holes, and the solution $\psi$ measures the deviation from a
  flat spacetime. Far away from the black holes one has $\psi\approx
  1$ with an almost Minkowski space, close to the holes we will find
  $\psi\sim 2$ with considerable curvature of spacetime.}. We solve
\begin{equation}\label{eq:BBH}
  \nabla^2\psi+\frac{1}{8}A^2\psi^{-7}=0
\end{equation}
for the function $\psi=\psi(x,y,z)$. $A^2=A^2(x,y,z)$ is a known,
positive function, and the computational domain is $\mathbbmss{R}^3$ with
two excised spheres,
\begin{equation}\label{eq:D-BBH}
  {\cal D}=\mathbbmss{R}^3-S_1-S_2.
\end{equation}
The radii $r_{1/2}$ and centers of the spheres are given.
The function $\psi$ must satisfy a Dirichlet boundary condition at
infinity, and Robin boundary conditions at the surface of each excised
sphere:
\begin{align}\label{eq:InftyBC-BBH}
  \psi\to 1&&&\mbox{as }r\to\infty\\
\label{eq:RobinBC-BBH}
  \dnachd{\psi}{r}+\frac{\psi}{2r_i}=0&&&\vec r\in\partial S_i,\;\; i=1,2
\end{align}
$\partial/\partial r$ in Eq.~\eqnref{eq:RobinBC-BBH} denotes the
radial derivative in a coordinate system centered at the center of
sphere $i$.

\begin{figure}
\begin{centering}
  \includegraphics[angle=90,scale=0.35]{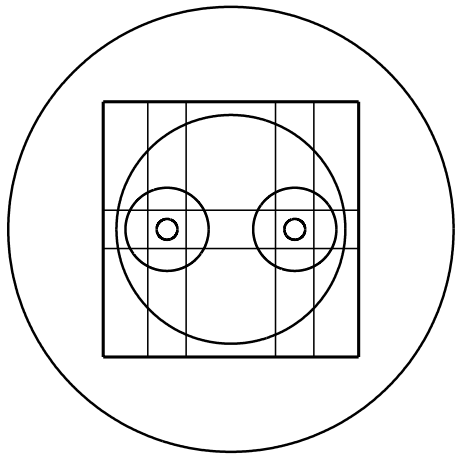}
\caption{\label{fig:Domains-BBH}
  Cut through the domain decomposition for the computational domain
  \eqnref{eq:D-BBH}.}
\end{centering}
\end{figure}
Figure \ref{fig:Domains-BBH} sketches the domain decomposition used
for the computational domain $\cal D$. We surround each excised sphere
with a spherical shell. These two spherical shells are matched
together with $5\times 3\times 3$ rectangular blocks, where the two
blocks that contain the excised spheres $S_{1,2}$ are removed.
Finally, we surround this structure with a third spherical shell
extending to very large outer radius. This gives a total of 46
subdomains, namely 3 shells and 43 rectangular blocks.

In the inner spheres we use a log mapping for the radial coordinate.
In the rectangular blocks, a combination of linear and logarithmic
mappings is used similar to the 2D example in figure
\ref{fig:SketchRectangles}. In the outer sphere an inverse mapping is
used which is well adapted to the fall-off behavior $\psi\sim
1+a\,r^{-1}+\cdots$ for large radii $r$. The outer radius of the outer
spherical shell is chosen to be $10^9$ or $10^{10}$ and a Dirichlet
boundary condition $\psi=1$ is used to approximate
Eq.~\eqnref{eq:InftyBC-BBH}.

We now present two solutions with different sizes and locations of the
excised spheres. In sections \ref{sec:Example2-Preconditioning} to
\ref{sec:Example2-ParallelExecution}, we then discuss several topics
including preconditioning and parallelization.

\subsubsection{Equal sized spheres}

First we choose two equal sized spheres with radii $r_1=r_2=1$.  The
separation between the centers of the spheres is chosen to be 10, the
outer radius of the outer sphere is $10^9$.

We solve equation \eqnref{eq:BBH} at several resolutions. The highest
resolution uses $29^3$ collocation points in each rectangular block,
$29\times 21\times 42$ collocation points (radial, $\theta$ and $\phi$
directions) in the inner spherical shells and $29\times 16\times 32$
in the outer spherical shell. We use the difference in the solutions
at neighboring resolutions as a measure of the error. We denote the
pointwise maximum of this difference by $L_{inf}$ and the
root-mean-square of the grid point values by $L_2$.  We also compute
at each resolution the quantity
\begin{equation}
\label{eq:M}
  M=-\frac{1}{2\pi}\int_{\infty}\dnachd{\psi}{r}d^2S
\end{equation}
which is the total mass of the binary black hole system. $M$ will be
needed in the comparison to a finite difference code below.  The
difference $\Delta M$ between $M$ at neighboring resolutions is again
a measure of the error of the solution.

\begin{figure}
\begin{centering}
  \includegraphics[scale=0.35]{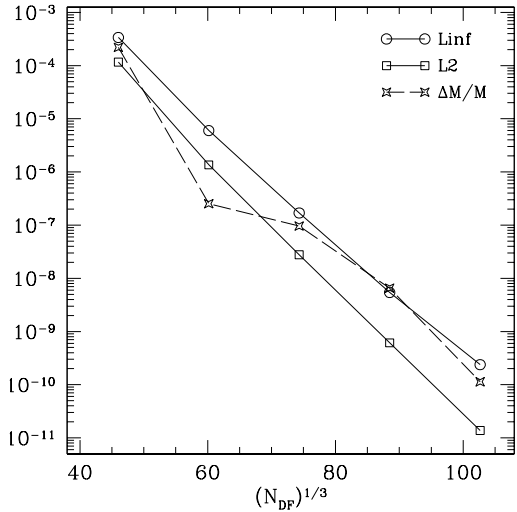}
\caption{\label{fig:Convergence-BBH}
  Convergence of solution of \eqnref{eq:BBH}-\eqnref{eq:RobinBC-BBH}
  with radii of excised spheres $r_1=r_2=1$ at separation $10$.
  $N_{DF}$, $L_{inf}$, $L_2$, $M$ and $\Delta M$ are defined in the
  text immediately before and after Eq.~\eqnref{eq:M}.}
\end{centering}
\end{figure}

Figure \ref{fig:Convergence-BBH} shows the convergence of the solution
$\psi$ with increasing resolution. Since the rectangular blocks and
the spheres have different numbers of collocation points, the cube
root of the total number of degrees of freedom, $N_{DF}^{1/3}$ is used
to label the $x$-axis.  The exponential convergence is apparent.
Because of the exponential convergence, and because Linf, L2 and
$\Delta M$ utilize differences to the next lower resolution, the
errors given in figure \ref{fig:Convergence-BBH} are essentially the
errors of the next {\em lower} resolution. Note that at the highest
resolutions the approximation of the outer boundary condition
{\eqnref{eq:InftyBC-BBH}} by a Dirichlet boundary condition at finite
outer radius $10^9$ becomes apparent: If we move the outer boundary to
$10^{10}$, $M$ changes by $2\cdot 10^{-9}$ which is of order $1/10^9$
as expected.

On the coarsest resolution $\psi=1$ is used as the initial guess.
Newton-Raphson then needs six iterations to converge. On the finer
resolutions we use the result of the previous level as the initial guess.
These initial guesses are so good that one Newton-Raphson iteration is
sufficient on each resolution.

\subsubsection{Nonequal spheres --- Different length scales}
\label{sec:Example2-LengthScales}

With the multidomain spectral method it is possible to distribute
resolution differently in each subdomain. This allows for geometries
with vastly different length scales. As an example, we again solve 
equations \eqnref{eq:BBH}-\eqnref{eq:RobinBC-BBH}. The radii of the
spheres are now $r_1=1$ and $r_2=0.05$, and the separation of the centers
of the spheres is $100$.  The separation of the holes is thus 2000
times the radius of the smaller sphere.  A finite difference code
based on a Cartesian grid for this geometry would have to use adaptive
mesh refinement.

\begin{figure}
\begin{centering}
  \includegraphics[scale=0.35]{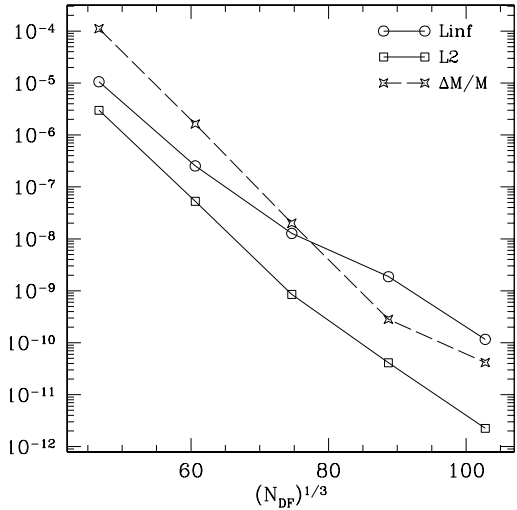}
\caption{\label{fig:Convergence-MassRatio}
  Convergence of solution of \eqnref{eq:BBH}-\eqnref{eq:RobinBC-BBH} for
  excised spheres of radii $r_1=1$, $r_2=0.05$ at separation $100$.
  Symbols as in figure \ref{fig:Convergence-BBH}.}
\end{centering}
\end{figure}

With the spectral solver, we still use the domain decomposition
depicted in figure \ref{fig:Domains-BBH}, but now the inner radii of
the two inner spherical shells are different. The outer boundary of
the outer sphere is at $10^{10}$.  The number of collocation points in
each subdomain is almost identical to the case with equal sized
spheres of figure \ref{fig:Convergence-BBH}, except we add 8
additional radial collocation points to the shell around the small
excised sphere.  As before we solve on different resolutions and
compute the norms of the differences of the solution between different
resolutions, as well as of the total mass $M$. The results are shown
in figure \ref{fig:Convergence-MassRatio}. The exponential convergence
shows that the solver can handle the different length scales involved
in this problem.

\subsubsection{Preconditioning}
\label{sec:Example2-Preconditioning}

The finite difference approximation ${\cal A}_{FD}$ is initialized
with the second derivative terms, the matching conditions in touching
domains [cf. Eqs.~\eqnref{eq:S3_D} and \eqnref{eq:S3_E}], and with a FD
approximation of the Robin boundary condition
Eq.~\eqnref{eq:RobinBC-BBH}.  Running on a single processor, we could
again define the preconditioner ${\cal B}$ via an ILU decomposition of
${\cal A}_{FD}$.  However, when running on multiple processors, an ILU
decomposition requires a prohibitive amount of communication, and
block ASM preconditioning\cite{Smith-Bjorstad-Gropp:1996} with one
block per processor becomes favorable.  After considerable
experimentation, we settled on an implicit definition of ${\cal B}$
via its action on vectors. ${\cal B}\underline{\bf v}$ is {\em
  defined} to be the approximate solution $\underline{\bf w}$ of
\begin{equation}\label{eq:PC-LinearSystem}
  {\cal A}_{FD}\underline{\bf w}=\underline{\bf v}.
\end{equation}

Equation \eqnref{eq:PC-LinearSystem} is solved using a second, inner
iterative solver, usually GMRES preconditioned with ILU (on a single
processor) or a block ASM method (on multiple processors).  The inner
solver is restricted to a fixed number of iterations.  Applying a
fixed number of iterations of an iterative solver is {\em not} a
linear operation, hence ${\cal B}$ represents no longer a matrix, but
a nonlinear operator.  In the outer linear solve we therefore use
FGMRES\cite{Saad:1993}, a variant of GMRES that does not require that
the preconditioner $\cal B$ is linear.  With this preconditioning the
outer linear solver needs about 20 iterations to reduce the residual
of the linear solve by $10^{-5}$.

More inner iterations reduce the number of iterations in the outer
linear solve, but increase the computations per outer iteration.  We
found the optimal number of inner iterations to be between 15-20.  In
all the computations given in this paper we use 20 inner iterations,
except for the 2-D example in table \ref{tab:Laplace2D} where 10 inner
iterations sufficed.

\subsubsection{Multigrid}

We also experimented with multigrid algorithms\cite{Canuto-Hussaini,
  NumericalRecipes} to improve the runtime.  The potential for
multigrid is fairly small, since the number of collocation points is
so low. In this particular problem, an accuracy of better than
$10^{-6}$ can be achieved with $17^3$ grid points per domain, which
limits multigrid to at most two coarse levels.

In addition it is not trivial to construct a restriction operator. The
obvious and canonical choice for a restriction operator is to
transform to spectral space, discard the upper half of the spectral
coefficients, and transform back to real space on a coarser grid. This
does not work here because the operator $\cal S$ uses the boundary
points of each subdomain to convey information about matching between
subdomains and about boundary conditions. Since these boundary points
are filled using different equations than the interior points, the
residual will typically be {\em discontinuous} between boundary points
of a subdomain and interior points.  Information about discontinuities
is mainly stored in the high frequency part of a spectral expansion
and discarding these will thus result in a loss of information about
matching between grids.  However, the coarse grid correction of a
multigrid algorithm is supposed to handle long wavelength modes of the
solution. In our case these extend over several subdomains and thus
information about matching is essential. Hence the simple approach of
discarding the upper half of the frequencies discards the most vital
parts of the information required by the coarse grid solver.

Thus one seems to be compelled to use a real space restriction
operator.  We examined straight injection\cite{NumericalRecipes}
which performed fairly well. The execution speed was comparable to the
preconditioning with an inner linear solve as described in section
\ref{sec:Example2-Preconditioning}. Since we did not achieve a
significant code speed-up, there was no reason to keep the increased
complexity of the multigrid algorithm.

\subsubsection{Comparison to a Finite Difference Code} 

The computational domain Eq.~\eqnref{eq:D-BBH} is challenging for 
standard finite difference codes based on a regular Cartesian grids
for two reasons:
\begin{enumerate}
\item The boundaries of the excised spheres do not coincide with
  coordinate boundaries, so complicated interpolation/extrapolation is
  needed to satisfy the boundary condition \eqnref{eq:RobinBC-BBH}
  (This problem led to a reformulation of the underlying physical
  problem without excised spheres\cite{Brandt-Brugmann:1997}).
\item Resolving both the rapid changes close to the excised spheres
  {\em and} the fall-off behavior toward infinity requires a large
  number of grid points.
\end{enumerate}

\begin{figure}
\begin{centering}
  \includegraphics[scale=0.4]{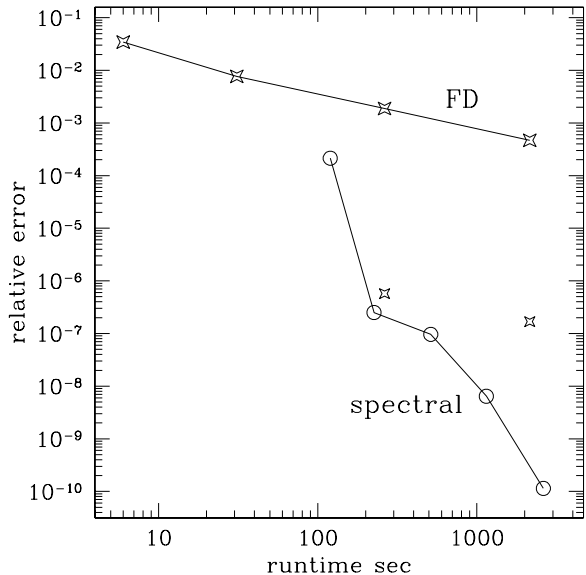}
\caption{\label{fig:CompareCadez}
  Comparison of runtime vs. achieved accuracy for the new spectral
  solver and the highly optimized Cad{\'e}z code. Plotted is the achieved
  accuracy of the total mass $M$ vs. runtime needed to solve Eqs.~\eqnref{eq:BBH}-\eqnref{eq:RobinBC-BBH} for both codes.}
\end{centering}
\end{figure}

In \cite{Cook-Choptuik-etal:1993} three different methods were
developed to solve the boundary value problem
\eqnref{eq:BBH}-\eqnref{eq:RobinBC-BBH}.  The best one turned out to
be a finite difference code based on a specially adapted coordinate
system, the so-called Cad{\'e}z coordinates. This code is an
FAS-multigrid algorithm developed specifically for the differential
equation \eqnref{eq:BBH}. Care was taken that the truncation error is
strictly even in grid-spacing $h$, thus allowing one to take two or
three solutions at different resolutions and Richardson extrapolate to
$h\to 0$.  The Cad{\'e}z code is thus specially built for this
equation in this geometry and it is unlikely that it can be
significantly improved upon by any finite difference method.

On the other hand, our spectral solver is general purpose. The domain
decomposition is not restricted to $\mathbbmss{R}^3$ with two excised
spheres and we do not employ any specific optimizations for this
particular problem.

We compare these two codes for the configuration with equal sized
spheres.  Figure \ref{fig:CompareCadez} shows a plot of runtime vs.
achieved accuracy for both codes.  These runs were performed on a
single RS6000 processor; the highest resolution runs needed about 1GB
of memory.  The solid line labeled FD represents the results of the
finite difference code without Richardson extrapolation. This line
converges quadratically in grid spacing.  The two stars represent
Richardson extrapolated values. The superiority of the spectral code
is obvious.  In accuracy, the spectral method outperforms even the
finite difference code with Richardson extrapolation by orders of
magnitude.  Only very few finite difference codes allow for Richardson
extrapolation, hence one should also compare the finite difference
code without Richardson extrapolation to the spectral code: Then the
{\em lowest} resolution of the spectral code is as accurate as the
{\em highest} resolution of the finite difference code and faster by a
factor of 20.  Note also that the Cad{\'e}z code cannot handle excised
spheres of very different sizes or spheres that are widely separated.
In particular, it cannot be used for the configuration in
section~\ref{sec:Example2-LengthScales}, which is readily solved by
our method.

\subsubsection{Parallelization}
\label{sec:Example2-ParallelExecution}

Most computations during a solve are local to each subdomain; the
operator $\cal S$ and the Jacobian $\cal J$ need communicate only
matching information across subdomains.  The inner linear solve is a
completely standard parallel linear solve with an explicitly known
matrix ${\cal A}_{FD}$. The software package PETSc has all the
necessary capabilities to solve this system of equations efficiently
in parallel. Hence parallelization by distributing different
subdomains to different processors is fairly straightforward.

However, different elements of the overall solve scale with different
powers of the number of collocation points per dimension.  If we
denote the number of collocation points per dimension by $N$, the
following scalings hold in three dimensions (the most interesting
case): A spectral transform in a rectangular domain requires
${\cal O}(N^3\log N)$ operations; the transform in a sphere ---where no
useful fast transform for the Legendre polynomials is available---
requires ${\cal O}(N^4)$ operations; interpolation to {\em one} point is
${\cal O}(N^3)$, so interpolation to {\em all} ${\cal O}(N^2)$ boundary points
scales like ${\cal O}(N^5)$. Thus the optimal assignment of subdomains to
processors is a function of $N$.  Moreover, assignment of subdomains to
processors is a discrete process --- it is not possible to move an
arbitrary fraction of computations from one processor to the another.
One always has to move a whole subdomain with all the computations
associated with it.  This makes efficient load balancing difficult.

At high resolution, the ${\cal O}(N^5)$ interpolation consumes most of
the runtime. Note that the outer spherical shell interpolates to $78$
block surfaces, whereas the inner shells each interpolate to $6$ block
surfaces. These interpolations are parallelized by first distributing
the data within each sphere to all processors. Then each processor
interpolates a fraction of the points and the results are gathered
again.

We present scaling results in table~\ref{tab:ParallelScaling}. These
results were obtained on the SP2 of the physics department of Wake
Forest University, and on NCSA's Platinum cluster, whose nodes have
two Intel Pentium processors each.  The listed times are cumulative
times for solving at five different resolutions, each solve using the
next lower solution as initial guess.  Not included in these times is
the set up in which the code determines which subdomain is responsible
for interpolating which ``overlapping'' boundary point. Also not
included is input/output.

\begin{table}
\begin{centering}
\caption{\label{tab:ParallelScaling}
  Runtime and scaling efficiency. Three processors host one shell and
$n_1$ blocks each, the remaining processors host $n_2$ blocks each. 
  The last four columns refer to the Platinum cluster.
}
\centerline{
\begin{tabular}{|c|cc|rc|rc|rc|}\hline
   & &
   & \multicolumn{2}{|c|}{SP2} 
   & \multicolumn{2}{|c|}{2 procs/node} 
   & \multicolumn{2}{|c|}{1 proc/node} \\
   Nprocs & $n_1$ & $n_2$ & t[sec] & eff. & t[sec] & eff. & t[sec] & eff.\\\hline
    1 &     &    & 2344 &      & 1654 &      & 1654 &      \\
    4 & 10  & 13 &  786 & 0.75 &  764 & 0.54 &  643 & 0.64 \\
    8 & 4-5 &  6 &  384 & 0.76 &  381 & 0.54 &  304 & 0.68 \\
   18 & 0   &  3 &      &      &  198 & 0.46 &  156 & 0.59 \\
   26 & 0   &  2 &      &      &  140 & 0.45 &  111 & 0.57 \\
   46 & 0   &  1 &      &      &   87 & 0.41 &   73 & 0.49 \\\hline
  \end{tabular}
}
\end{centering}
\end{table}

On the SP2 we achieve a scaling efficiency of 75\%, whereas the Intel
cluster has a lower scaling efficiency between around 54\% (8
processors), and 41\% (46 processors).  Given all the limitations
mentioned above these numbers are very encouraging.

Changing from serial to parallel execution degrades performance in two
ways: First, the ILU preconditioner used within the approximate inner
linear solve is replaced by an overlapping block ASM preconditioner.
Since this preconditioner is less efficient than ILU, the approximate
inner linear solve is less accurate after its fixed number of
iterations. Therefore the outer linear solve needs more iterations to
converge to the required accuracy of $10^{-5}$. The single processor
code needs 19 outer linear iterations, whereas the parallel codes need
23 or 24.  Thus the maximally achievable scaling efficiency is
limited to $19/23\approx 0.83$. The scaling efficiency on the SP2 is
close to this limit.

The second reason for low scaling efficiency is that we have not
optimized the MPI calls in any way. The fact that the scaling
efficiency on the cluster is much better if only one processor per node
is used, suggests that the MPI calls are a bottleneck. Using both
processors on a node doubles the communication load on that node which
doubles the waiting time for MPI communication. The higher scaling
efficiency on the SP2 which has faster switches also suggests that
the runs on the PC cluster are communication limited.

\subsection{Coupled PDEs in nonflat geometry with excised spheres}
\label{sec:Example3}

So far we have been considering only PDEs in a single variable.
However, the definition of the operator $\cal S$ is not restricted to
this case. In this section we present a solution of four coupled
nonlinear PDEs.  These equations are

\begin{figure}
\begin{centering}
  \includegraphics[scale=0.35]{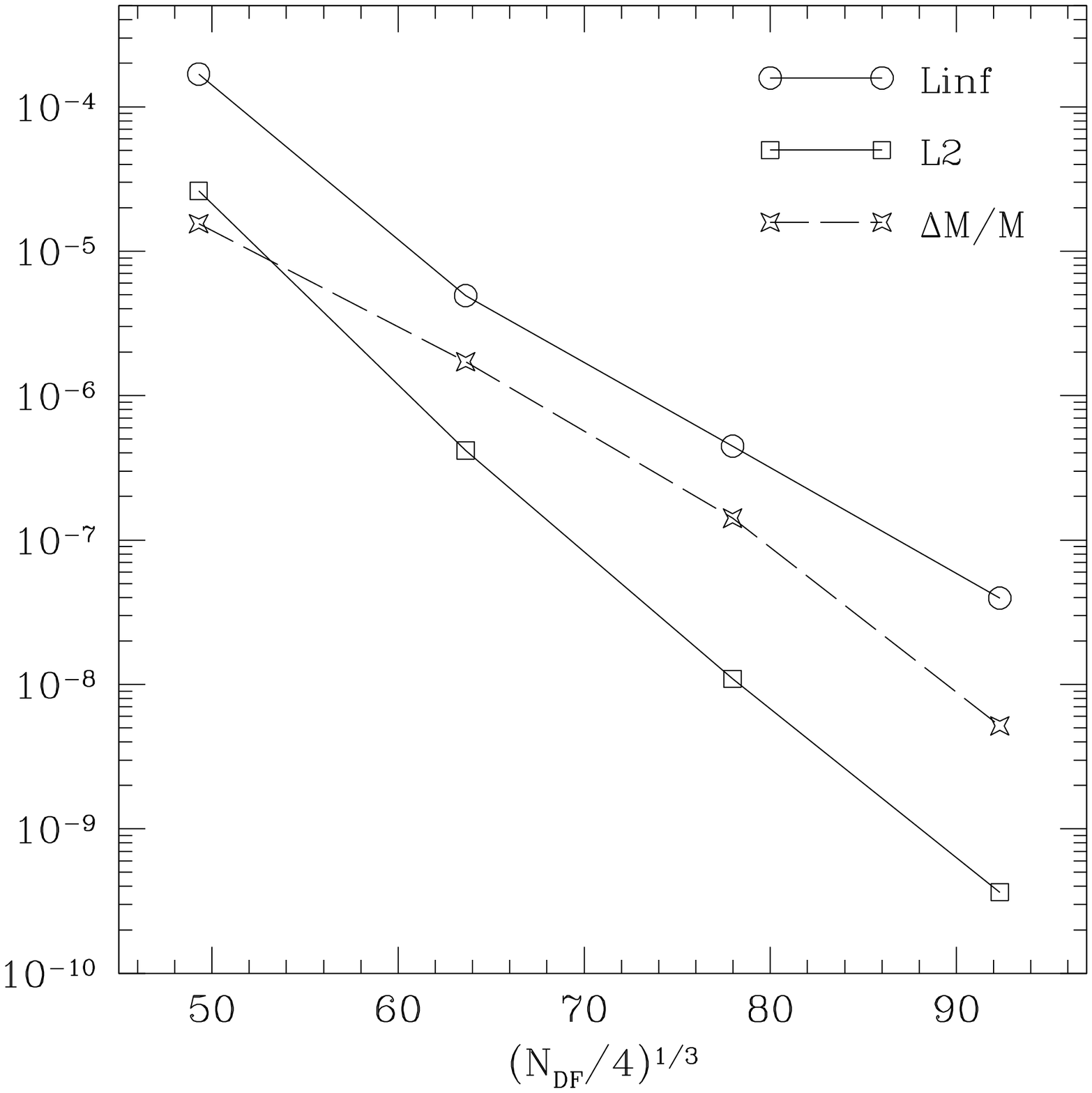}
\caption{\label{fig:Convergence-ConfTT}
  Convergence of solution to coupled PDEs \eqnref{eq:ConfTT-1} and
  \eqnref{eq:ConfTT-2}.  Definitions as in figure \ref{fig:Convergence-BBH}.}
\end{centering}
\end{figure}

\begin{align}
\label{eq:ConfTT-1}
    &\tilde\nabla^2\psi-\frac{1}{8}\psi\tilde R-\frac{1}{12}\psi^5K^2
    +\frac{1}{8}\psi^{-7}\sum_{i,j=1}^3\tilde A_{ij}\tilde A^{ij}=0,\\
\label{eq:ConfTT-2}
    &\qquad\tildeLapLong V^i-\frac{2}{3}\psi^6\tilde\nabla^iK
    +\sum_{j=1}^3\tilde\nabla_{\!j}\tilde M^{ij}=0,&\hspace*{-2cm}i=1,2,3
\end{align}
These equations are important for the binary black hole problem.  The
exact definitions of the various terms can be found in
\cite{Pfeiffer-Teukolsky-Cook:2001}.  For this paper, only the
following information is necessary: $\tilde\nabla^2$ is the Laplace
operator on a nonflat three-dimensional manifold, hence
Eq.~\eqnref{eq:ConfTT-1} is an elliptic equation for $\psi$.
$\tildeLapLong$ is a variant of the vector Laplacian, thus
Eq.~\eqnref{eq:ConfTT-2} is an elliptic equation for the vector $V^i,
i=1,2,3$. The variables $\tilde A_{ij}$ and $\tilde A^{ij}$ are
functions of $V^i$, so that Eqs.~\eqnref{eq:ConfTT-1} and
\eqnref{eq:ConfTT-2} have to be solved simultaneously. The functions
$\tilde R$, $K$ and $\tilde M^{ij}$ are given.

The computational domain again has two spheres $S_{1,2}$ excised, 
\begin{equation}
  {\cal D}=\mathbbmss{R}^3-S_1-S_2.
\end{equation}
The radii of the excised spheres are $r_1=r_2=2$ and the separation
between the centers is 10.

We have Dirichlet boundary conditions on all boundaries:
\begin{align}
  \psi&=1,\\
  V^i&=0,\qquad i=1,2,3.
\end{align}

We solve the problem again at several different resolutions. On the
coarsest level two Newton-Raphson iterations are necessary, whereas
the finer levels need only one Newton-Raphson iteration. The linear
solver needs 30 iterations to reduce the residual by $10^5$.  Details
about constructing ${\cal A}_{FD}$ for the nonflat differential
operators $\tilde\nabla^2$ and $\tildeLapLong$ are given in appendix
\ref{sec:Nonflat-Preconditioning}.

The convergence of the solutions is shown in figure
\ref{fig:Convergence-ConfTT}. We again find smooth exponential
convergence. Recall that the plotted quantities essentially give the
error of the next lower resolution. Hence the next-to-highest
resolution run with a total of $78^3\approx 500000$ collocation points
has a maximum pointwise error of $\sim 0.5\cdot 10^{-7}$. The wall
clock time for that run is less than 2 hours on four RS 6000
processors.

This problem has also been attacked with a finite difference
code\cite{Marronetti-Matzner:2000}.  The finite difference code required a
runtime of 200 CPU hours (on 16 nodes of an Origin 2000). The accuracy
of the finite difference code seems to be comparable to the lowest
resolution solve of our spectral solver, which took 16 minutes CPU
time.  Compared to the finite difference code the spectral code is
almost embarrassingly fast.

\section{Improvements}

The fact that spherical harmonics have fewer spectral coefficients
than collocation points causes a host of complications. We have to
solve for mixed real-spectral coefficients,
Eq.~\eqnref{eq:ExpansionSphere2}. This complicates the operator $\cal
S$, and severely complicates real space finite difference
preconditioning. A double Fourier series\cite{Orszag:1974} for the
angular variables might be superior to the expansion in spherical
harmonics, since this avoids the necessity for mixed real-spectral
coefficients. Moreover one can then use fast transforms for both
$\phi$ and $\theta$ which might improve runtime in the spherical
shells.

We are working on cylinders as a third possible subdomain type.  We
also hope to use more general mappings that are no longer restricted
to acting on each dimension separately.

In terms of pure runtime, one should try to optimize the interpolation
to boundary points of overlapping subdomains. This is the part of the
code that has the worst scaling with the number of unknowns. Replacing
the explicit summation of the series by one of the faster methods
discussed in \cite{Boyd:2001} should speed up the code tremendously.
As was seen in the discussion of parallelization in section
\ref{sec:Example2-ParallelExecution}, the code seems to be communication
limited on a PC cluster. One should therefore also optimize the MPI
calls. For example, one could overlap communication with subdomain
internal computations.

Even without all these improvements our code is already fairly fast.
This indicates the potential of our approach.

\section{Conclusion}

We have presented a new elliptic solver based on pseudo-spectral
collocation.  The solver uses domain decomposition with spherical
shells and rectangular blocks, and can handle nonlinear coupled
partial differential equations.

We introduced a new method to combine the differential operator, the
boundary conditions and matching between subdomains in {\em one}
operator $\cal S$. The equation ${\cal S}\underline{\bf u}=0$ is then
solved with Newton-Raphson and an iterative linear solver.  We show
than one can employ standard software packages for nonlinear and
linear solves and for preconditioning.

The operator $\cal S$ has the added benefit that it is modular.
Therefore adaption of the method to a new PDE or to new boundary
conditions is easy; the user has only to code the differential
operator and/or the new boundary conditions.  We also discuss our
treatment of mappings which decouples mappings from the actual code
evaluating the differential operator, and from the code dealing with
basis functions and details of spectral transforms. This modularity
again simplifies extension of the existing code with e.g. new
mappings.

We demonstrated the capabilities of the new method with three examples
on non-simply-connected computational domains in two and three
dimensions and with one and four variables.  We also demonstrated that
the domain decomposition allows for vastly different length scales in
the problem.  During the examples we discussed various practical
details like preconditioning and parallelization.  Two of these
examples were real applications from numerical relativity.  We found
the spectral code at the {\em coarsest} resolution to be as accurate
as finite difference methods, but faster by one to two orders of
magnitude. 

\appendix
\section{Preconditioning of inverse mappings}
\label{sec:FD-details}

In a subdomain with inverse mapping that extends out to (almost)
infinity, the outermost grid points are distributed very unevenly in
physical space.  This causes finite-difference approximations of
derivatives to fail if they are based on the physical coordinate
positions.  Therefore we difference in the collocation coordinate $X$
and apply the mapping via Eq.~\eqnref{eq:SecondDeriv-Mapped}. At the
collocation grid point $X_i$ with grid spacing $h_-=X_i-X_{i-1}$ and
$h_+=X_{i+1}-X_i$ we thus use
\begin{align}
\label{eq:FD-dudX}
\left(\dnachd{u}{X}\right)_{\!i}
=&-\frac{h_+u_{i-1}}{h_-(h_-+h_+)}
+\frac{(h_+-h_-)u_i}{h_+h_-}
+\frac{h_-u_{i+1}}{h_+(h_-+h_+)},\\
\label{eq:FD-du2dX2}
\left(\dnachd{^2u}{X^2}\right)_{\!i}
=& \frac{2u_{i-1}}{h_-(h_-+h_+)}
  -\frac{2u_i}{h_-h_+}
  +\frac{2u_{i+1}}{h_+(h_-+h_+)},\\
\label{eq:FD-du2dx2}
\left(\dnachd{^2u}{x^2}\right)_{\!i}
=&\;{X'_i}^2\left(\dnachd{^2u}{X^2}\right)_{\!i}
  +X^{\prime\prime}_i\left(\dnachd{u}{X}\right)_{\!i}.
\end{align}
If one substitutes Eqs.~\eqnref{eq:FD-dudX} and \eqnref{eq:FD-du2dX2} into
\eqnref{eq:FD-du2dx2}, then the coefficients of $u_{i-1}, u_i$ and
$u_{i+1}$ are the values that have to be entered into the
FD-approximation matrix ${\cal A}_{FD}$.

Even with this trick, preconditioning of the radial derivatives in an
extremely stretched outer sphere is not yet sufficiently good.  The
preconditioned Jacobian ${\cal BJ}$ still contains eigenvalues of size
$\sim 40$. The eigenmodes are associated with the highest radial mode
in the outer sphere. We found that we can suppress these eigenvalues
by damping this highest radial mode by a factor of 10 after the PETSc
preconditioning is applied.

\section{Preconditioning the nonflat Laplacian}
\label{sec:Nonflat-Preconditioning}

In a nonflat manifold, the Laplace operator of Eq.~\eqnref{eq:ConfTT-1}
contains second and first derivatives of $\psi$,
\begin{equation}\label{eq:nonflat-Laplace}
  \tilde\nabla^2\psi
=\sum_{i,j=1}^3g^{ij}\frac{\partial^2\psi}{\partial x^i\partial x^j}
 +\sum_{i=1}^3f^i\dnachd{\psi}{x^i}.
\end{equation}
The coefficients $g^{ij}$ and $f^i$ are known functions of position.
Since our particular manifold is almost flat, we have $g^{ii}\approx
1$, and $g^{ij}\approx 0$ for $i\neq j$.  We base our
preconditioning only on the diagonal part of \eqnref{eq:nonflat-Laplace}, 

\begin{equation}\label{eq:nonflat-Laplacian-diagonal}
\sum_{i=1}^3g^{ii}\frac{\partial^2}{\partial {x^i}^2}.
\end{equation}

In rectangular blocks, Eq.~\eqnref{eq:nonflat-Laplacian-diagonal} can
be preconditioned without additional fill-in in ${\cal A}_{FD}$.
Inclusion of the mixed second derivatives from
Eq.~\eqnref{eq:nonflat-Laplace} in ${\cal A}_{FD}$ leads to a large
fill-in of ${\cal A}_{FD}$. The increase in computation time due to
the larger fill-in outweighs the improvement of convergence of the
iterative solver in our problems.

For the spherical shells, matters are complicated by the fact that we
use mixed real-space/spectral space coefficients [recall
Eq.~\eqnref{eq:ExpansionSphere2}].  It is easy to precondition the
angular piece of the flat space Laplacian, since our basis functions
$Y_{lm}$ are the eigenfunctions of this operator.  Derivative
operators with angle-dependent coefficients lead to convolutions in
spectral space and lead thus to a large fill-in in the preconditioning
matrix.  Therefore we can only precondition radial derivatives with
coefficients that are {\em independent} of the angles $\theta, \phi$.
We thus need to approximate Eq.~\eqnref{eq:nonflat-Laplacian-diagonal}
by a flat space angular Laplacian and constant coefficient radial
derivatives. 
We proceed as follows.

Rewrite Eq.~\eqnref{eq:nonflat-Laplacian-diagonal} in spherical
coordinates,
\begin{align}
\sum_{i=1}^3g^{ii} \frac{\partial^2}{\partial{x^i}^2}
=\;&G^{\theta\theta}\dnachd{^2}{\theta^2}
  +G^{\phi\phi}\frac{\partial^2}{\partial\phi^2}
  +G^{rr}\frac{\partial^2}{\partial r^2}
+G^{\theta\phi}\dnachd{^2}{\theta\partial\phi}\nonumber\\\label{eq:Cart2Sph}
  &+G^{\theta r}\frac{\partial^2}{\partial\theta\partial r}
  +G^{\phi r}\frac{\partial^2}{\partial\phi\partial r}
+F^{\theta}\dnachd{}{\theta}
  +F^{\phi}\dnachd{}{\phi}
  +F^{r}\dnachd{}{r}.
\end{align}
At each grid point, the various functions $G$ and $F$ can be computed
via standard Cartesian to polar coordinate transformations.

For each radial grid point $r_i$, average over the angles to obtain
$\bar G^{\theta\theta}_i$, $\bar G^{rr}_i$ and $\bar F^r_i$.  Now
precondition as if $\bar G^{\theta\theta}_i$ were part of an angular
piece of the flat space Laplacian, i.e. enter $-l(l+1)\bar
G^{\theta\theta}_i/r_i^2$ as the diagonal element belonging to the
unknown $\hat u_{ilm}$.  Further, precondition $\bar
G^{rr}_i\partial^2/\partial r^2+\bar F^r_i\partial/\partial r$ with
finite differences as described in appendix \ref{sec:FD-details}.
Ignore all other terms in Eq.~\eqnref{eq:Cart2Sph}.

The operator $\tildeLapLong$ in Eq.~\eqnref{eq:ConfTT-2} is defined by
\begin{equation}
  \tildeLapLong V^i
\equiv
\tilde\nabla^2 V^i+\frac{1}{3}\sum_{k=1}^3\tilde\nabla^i \tilde\nabla_kV^k
+\sum_{k=1}^3\tilde R^i_kV^k, 
\end{equation}
$\tilde\nabla$ and $\tilde R_{ij}$ being the covariant derivative
operator and Ricci tensor associated with the metric of the manifold.
$\tildeLapLong V^i$ contains thus the nonflat Laplace operator acting
on each component $V^i$ separately, plus terms coupling the
different components which involve second derivatives, too.  We
precondition only the Laplace operator $\tilde\nabla^2 V^i$ for each
component $V^i$ separately as described above and ignore the coupling
terms between different components.

\begin{acknowledge}
  We thank Gregory Cook for helpful discussions.  This work was
  supported in part by NSF grants PHY-9800737 and PHY-9900672 to
  Cornell University. Computations were performed on the IBM SP2 of
  the Department of Physics, Wake Forest University, with support from
  an IBM SUR grant, as well as on the Platinum cluster of NCSA.

\end{acknowledge}

\bibliographystyle{abbrv}
\bibliography{Pfeiffer}

\end{document}